\documentclass[amsmath,prd,showpacs,preprintnumbers,nofootinbib,superscriptaddress]{revtex4-1}
\pdfoutput=1
\usepackage{graphicx}
\usepackage{dcolumn}
\usepackage{epsfig}
\usepackage{epstopdf}
\usepackage{epsf}
\usepackage{amssymb}
\usepackage{graphicx}
\usepackage{caption}
\usepackage{color}
\usepackage{subcaption}
\def\be{\begin{equation}}
\def\ee{\end{equation}}
\def\ba{\begin{eqnarray}}
\def\ea{\end{eqnarray}}

\begin{document}
\preprint{PI/UAN-2012-553FT}
\title{
The different varieties of the Suyama-Yamaguchi consistency relation and its violation as a signal of statistical inhomogeneity}

\author{Yeinzon Rodr\'iguez}\email{yeinzon.rodriguez@uan.edu.co}
\affiliation{Centro de Investigaciones, Universidad Antonio Nari\~no,  Cra 3 Este \# 47A-15, Bogot\'a D.C. 110231, Colombia}
\affiliation{Yukawa Institute for Theoretical Physics, Kyoto University, Kitashirakawa-Oiwake-cho, Sakyo-ku, Kyoto 606-8502, Japan}
\affiliation{Escuela de F\'isica, Universidad Industrial de Santander,  Ciudad Universitaria, Bucaramanga 680002, Colombia}
\author{Juan P. Beltr\'an Almeida}\email{juanpbeltran@uan.edu.co}
\affiliation{Centro de Investigaciones, Universidad Antonio Nari\~no,  Cra 3 Este \# 47A-15, Bogot\'a D.C. 110231, Colombia}
\author{C\'esar A. Valenzuela-Toledo}\email{cesar.valenzuela@correounivalle.edu.co}
\affiliation{Departamento de F\'isica, Universidad del Valle, Ciudad Universitaria Mel\'endez, Santiago de Cali 760032, Colombia}


\begin{abstract}
We present the different consistency relations that can be seen as variations of the well known Suyama-Yamaguchi (SY) consistency relation  $\tau_{\rm NL} \geqslant \left(\frac{6}{5} f_{\rm NL} \right)^2$, the latter involving the levels of non-gaussianity $f_{\rm NL}$ and $\tau_{\rm NL}$ in the primordial curvature perturbation $\zeta$. It has been (implicitly) claimed that the following variation:
$\tau_{\rm NL} ({\bf k}_1, {\bf k}_3) \geqslant \left(\frac{6}{5}\right)^2 f_{\rm NL} ({\bf k}_1) f_{\rm NL} ({\bf k}_3)$, which we call ``the fourth variety'',
in the collapsed (for $\tau_{\rm NL}$) and squeezed (for $f_{\rm NL}$) limits is always satisfied independently of any physics;  however, the proof depends sensitively on the assumption of scale-invariance 
(expressing this way the fourth variety of the SY consistency relation as $\tau_{\rm NL} \geqslant \left(\frac{6}{5} f_{\rm NL} \right)^2$)
which only applies for cosmological models involving Lorentz-invariant scalar fields (at least at tree level), leaving room for a strong violation of this variety of the consistency relation when non-trivial degrees of freedom, for instance vector fields, are in charge of the generation of the primordial curvature perturbation.  
With this in mind as a motivation, we explicitly state, in the first part of this work, under which conditions the SY consistency relation has been claimed to hold in its different varieties (implicitly) presented in the literature since its inception back in 2008; as a result, we show for the first time that the variety $\tau_{\rm NL} ({\bf k}_1, {\bf k}_1) \geqslant \left(\frac{6}{5} f_{\rm NL} ({\bf k}_1) \right)^2$, which we call ``the fifth variety'', is always satisfied even when there is strong scale-dependence and high levels of statistical anisotropy as long as statistical homogeneity holds: thus, an observed violation of this specific variety would prevent the comparison between theory and observation, shaking this way the foundations of cosmology as a science.
In the second part, we concern about the existence of non-trivial degrees of freedom, concretely vector fields for which the levels of non-gaussianity have been calculated for very few models; among them, and by making use of the $\delta N$ formalism at tree level,  
we study a class of models that includes the vector curvaton scenario, vector inflation, and the hybrid inflation with coupled vector and scalar ``waterfall field'' where $\zeta$ is generated at the end of inflation,
finding that the fourth variety of the SY consistency relation is indeed strongly  violated for some specific wavevector configurations while the fifth variety continues to be well satisfied.
Finally, as a byproduct of our investigation, we draw attention to a quite recently demonstrated variety of the SY consistency relation: $\tau^{\rm iso}_{\rm NL}\geqslant(\frac{6}{5}f^{\rm iso}_{\rm NL})^2$, in scenarios where scalar and vector fields contribute to the generation of the primordial curvature perturbation;  this variety of the SY consistency relation is satisfied although the isotropic pieces of the non-gaussianity parameters receive contributions from the vector fields. We discuss further implications for observational cosmology. 
\end{abstract}

\pacs{98.80.Cq}

\maketitle

\section{Introduction: The first variety of the SY consistency relation} \label{first}

Modern cosmology not only studies the background dynamics of any inflationary model of the Universe, but also the perturbation dynamics via the connected $n$-point correlators $\langle \zeta({\bf k}_1) \zeta({\bf k}_2) ... \zeta({\bf k}_n) \rangle_c$ of the primordial curvature perturbation $\zeta$ \cite{lythbook}.  During some time, the two-point correlator of $\zeta$ was enough for the purpose of comparing theoretical predictions with observation, but, after satellite missions reached amazing accuracy levels \cite{wmap, planck} (see for example the recent results by Planck \cite{Planck2013XVI, Planck2013XXII, Planck2013XXIII, Planck2013XXIV}), it has been necessary to work out the connected three- and four-point correlators of $\zeta$.  Non-vanishing connected three- or four-point correlators of $\zeta$ imply that the probability distribution function of the primordial curvature perturbation is non-gaussian \cite{ValenzuelaToledo:2011fj}, that being the reason why the functions $f_{\rm NL} ({\bf k}_1, {\bf k}_2, {\bf k}_3)$, and $\tau_{\rm NL} ({\bf k}_1, {\bf k}_2, {\bf k}_3, {\bf k}_4)$ and $g_{\rm NL} ({\bf k}_1, {\bf k}_2, {\bf k}_3, {\bf k}_4)$, that parameterize the connected three- and four-point correlators of $\zeta$ respectively are called the levels of non-gaussianity.  

A few years ago, Suyama and Yamaguchi \cite{Suyama:2007bg} showed that $f_{\rm NL}$ and $\tau_{\rm NL}$ satisfy the following consistency relation:
\begin{equation}
\boxed{\tau_{\rm NL} \geqslant \left( \frac{6}{5} f_{\rm NL} \right)^2} \,,  \label{sye}
\end{equation}
as long as the following conditions are satisfied:
\begin{itemize}
\item {\it Condition 1}: The calculation of $f_{\rm NL}$ and $\tau_{\rm NL}$ is performed at tree level in the diagrammatic approach of the $\delta N$ formalism \cite{bksw,ValenzuelaToledo:2011fj}.
\item {\it Condition 2}: The inflationary dynamics is driven by any number of slowly-rolling scalar fields.
\item {\it Condition 3}: The fields involved are gaussian.
\item {\it Condition 4}: The field perturbations are scale-invariant.
\end{itemize}
Although $f_{\rm NL}$ and $\tau_{\rm NL}$ are not directly comparable in the general case because they are functions of the wavevectors, the latter expression is valid since, under the conditions previously stated, $f_{\rm NL}$ and $\tau_{\rm NL}$ are scale-invariant \cite{Lyth:2005fi}. We will call Eq. (\ref{sye}) the first variety of the Suyama-Yamaguchi (SY) consistency relation.

Because of the role of the SY consistency relation of Eq. (\ref{sye}) at ruling out classes of cosmological inflationary models in the event that the inequality is observationally violated, it is reasonable to wonder if it is possible to relax some or all the conditions previously stated while preserving the SY consistency relation.  Suyama, Takahashi, Yamaguchi, and Yokoyama \cite{styy}, and Sugiyama, Komatsu, and Futamase \cite{Sugiyama:2011jt}, made efforts in this direction, trying to take into account the one-loop corrections to the three- and four-point correlation functions (relaxing condition 1), but adding a new condition:
\begin{itemize}
\item {\it Condition 5}: $f_{\rm NL} ({\bf k}_1, {\bf k}_2, {\bf k}_3)$ is evaluated in the squeezed limit (${\bf k}_1 \rightarrow 0$) while $\tau_{\rm NL} ({\bf k}_1, {\bf k}_2, {\bf k}_3, {\bf k}_4)$ is evaluated in the collapsed limit (${\bf k}_1 + {\bf k}_2 \rightarrow 0$).
\end{itemize}
This effort was consolidated by Sugiyama \cite{sal} who demonstrated that the SY consistency relation is valid if conditions 2-5 are met and condition 1 is replaced by the following one:
\begin{itemize}
\item {\it Condition 1}: The calculation of $f_{\rm NL}$ and $\tau_{\rm NL}$ is performed at all loop corrections in the diagrammatic approach of the $\delta N$ formalism \cite{bksw,ValenzuelaToledo:2011fj}.
\end{itemize}
However, in an elegant work by Smith, LoVerde, and Zaldarriaga \cite{cr2}, it was shown that the first variety of the SY consistency relation is valid if condition 5 is met while conditions 1-4 are replaced by the following ones:
\begin{itemize}
\item {\it Condition 1}: The calculation of $f_{\rm NL}$ and $\tau_{\rm NL}$ is performed non-perturbatively.
\item {\it Condition 2}: The inflationary dynamics is arbitrary.
\item {\it Condition 2a (actually, a consequence of condition 2)}: The fields involved (if any) can be non-gaussian.
\item {\it Condition 3}: Statistical homogeneity of $\zeta$ is preserved.
\item {\it Condition 4}: Statistical isotropy of $\zeta$ is preserved.
\end{itemize}


The statistical homogeneity of $\zeta$ states that the connected $n$-point correlators in the configuration space are invariant under space translations \cite{ValenzuelaToledo:2011fj};  this implies that, in the momentum space, the connected $n$-point correlators are proportional to a Dirac delta function:
\begin{equation}
\langle \zeta({\bf k}_1) \zeta({\bf k}_2) ... \zeta({\bf k}_n) \rangle_c \propto \delta^3 ({\bf k}_1 + {\bf k}_2 + ... + {\bf k}_n) \,.
\end{equation}
The above implies that the wavevector configuration is such that ${\bf k}_1$, ${\bf k}_2$, ..., and ${\bf k}_n$ form a warped polygon.  The statistical homogeneity is fundamental in cosmology since it allows us to compare theoretical predictions with observations (via the ergodic theorem, see Refs. \cite{ValenzuelaToledo:2011fj,weinbergbook}).  In contrast, the statistical isotropy of $\zeta$ states that the connected $n$-point correlators in the configuration space are invariant under space rotations \cite{ValenzuelaToledo:2011fj};  this implies that, in the momentum space, and once statistical homogeneity has been imposed, each connected $n$-point correlator is proportional to a function $M_\zeta$ called the $n-1$-spectrum which is invariant under rotations in the momentum space:
\begin{equation}
\langle \zeta({\bf k}_1) \zeta({\bf k}_2) ... \zeta({\bf k}_n) \rangle_c = (2\pi)^3 \delta^3 ({\bf k}_1 + {\bf k}_2 + ... + {\bf k}_n)  M_\zeta ({\bf k}_1, {\bf k}_2, ... , {\bf k}_n) \,,
\end{equation}
where
\begin{equation}
M_\zeta (\tilde{\bf k}_1, \tilde{\bf k}_2, ... ,\tilde{\bf k}_n) = M_\zeta ({\bf k}_1, {\bf k}_2, ... , {\bf k}_n) \,,  \label{spectra}
\end{equation}
being $\tilde{\bf k}_i = \mathcal{R} \ {\bf k}_i$ with $\mathcal{R}$ being the rotation operator such that $\tilde{\bf x}_i = \mathcal{R} \ {\bf x}_i$ in the configuration space, i.e. the $n-1$-spectrum is the same no matter the orientation of the warped polygon.  The statistical isotropy is the usual assumption in cosmology since the inflationary dynamics is usually driven by scalar fields only, which, due to its nature, do not exhibit preferred directions.

Some time after the work by Smith, LoVerde, and Zaldarriaga, and by employing an alternative technique, Assassi, Baumann, and Green \cite{abg} obtained the same result as the former ones under the same assumptions.  However, both proofs rely on this additional assumption:
\begin{itemize}
\item {\it Condition 6}: $f_{\rm NL}$ and $\tau_{\rm NL}$ are scale-invariant,
\end{itemize}
as clearly stated in a later work by Kehagias and Riotto \cite{kr}.  The latter authors come back to the original conditions 1 and 5, except for condition 2 which is reinforced and conditions 3 and 4 which are relaxed:
\begin{itemize}
\item {\it Condition 2}: The inflationary dynamics is driven by any number of slowly-rolling Lorentz-invariant scalar fields.
\item {\it Condition 3}: The fields involved can be non-gaussian.
\item {\it Condition 4}: The field perturbations, in principle, might not be scale-invariant.
\end{itemize}
By paying attention to the symmetries present during the de Sitter epoch, and employing the operator product expansion technique, Kehagias and Riotto show that the connected two-, three-, and four-point correlators of the field perturbations are scale-invariant in the squeezed (for the three-point correlators) and collapsed (for the four-point correlators) limits.  Thus, the first variety of the SY consistency relation, under the conditions just stated above, is shown to be valid but just at tree level in the diagrammatic approach to the $\delta N$ formalism.  Kehagias and Riotto assume that their proof is valid, even including loop corrections, for a wide range of models since loop corrections seem to become important just for rather marginal cases \cite{Tasinato:2012js} (see however \cite{Cogollo:2008bi,Rodriguez:2008hy});  nevertheless, their proof at all loop corrections, employing the results by Sugiyama \cite{sal}, is still incomplete as long as the connected $n$-point correlators in the field perturbations, with $n > 4$, are not shown to be scale-invariant in the required limits.

The proofs by Smith, LoVerde, and Zaldarriaga \cite{cr2}, and by Assassi, Baumann, and Green \cite{abg}, are by far the most general and lead us to wonder whether one or some of the conditions these proofs are based on are unnecessary;  they also lead us to state, assuming that the 
proof in \cite{kr} is valid at all orders, that a violation of the first variety of the SY consistency relation would imply that new non-trivial degrees of freedom play a role during inflation.  Among the conditions that could be unnecessary, we think that the statistical isotropy condition, and the scale-invariance condition for $f_{\rm NL}$ and $\tau_{\rm NL}$, are the ones to investigate since the statistical homogeneity condition is essential to compare theory with observation \cite{ValenzuelaToledo:2011fj,weinbergbook} and since the other three conditions are absolutely general.  Among the possible new non-trivial degrees of freedom, the vector fields are particularly interesting since they are suitable candidates to explain the apparent violation of statistical isotropy observed in the 5-year data from the WMAP satellite \cite{gr,Groeneboom:2008fz} and the anomalies at low multipoles confirmed by the 1-year data from the Planck satellite \cite{Planck2013XXIII}, given that they define inherently a preferred direction for the expansion, for the distribution of primordial fluctuations, or for both \cite{ValenzuelaToledo:2011fj}, \cite{vi}-\cite{dklr}.   The vector fields, the statistical anisotropy, and the scale-dependence of $f_{\rm NL}$ and $\tau_{\rm NL}$ are indeed related to each other since, even in the simplest case when the expansion is isotropic, vector fields break the statistical isotropy in $\zeta$ \cite{dklr}, and can make $f_{\rm NL}$ and $\tau_{\rm NL}$ strongly scale-dependent at tree level \cite{varios} and even more at higher levels in the perturbative expansion \cite{ValenzuelaToledo:2011fj}. Several interesting works have recently studied this relation between non-gaussianity and statistical anisotropy in different models, see for instance Refs. \cite{varios}-\cite{Lyth:2013sha}.



\section{Relaxing the statistical isotropy condition: the second and third varieties of the SY consistency relation} \label{second}

In this section, we are going to reproduce the proof by Smith, LoVerde, and Zaldarriaga \cite{cr2}, step by step, but this time relaxing the statistical isotropy condition.  As discussed around Eq. (\ref{spectra}), after imposing statistical homogeneity, the connected $n$-point correlator of $\zeta$ is written in terms of a function $M_\zeta ({\bf k}_1, {\bf k}_2, ..., {\bf k}_n)$, called the $n-1$-spectrum, which, if statistical isotropy is imposed, becomes invariant under spatial rotations of its arguments; this in turn implies that the spectrum $P_\zeta ({\bf k}_1, {\bf k}_2)$ and bispectrum $B_\zeta ({\bf k}_1, {\bf k}_2, {\bf k}_3)$ depend in this case only on the magnitudes of the wavevectors involved (i.e., $P_\zeta ({\bf k}_1, {\bf k}_2) = P_\zeta (k)$ and $B_\zeta ({\bf k}_1, {\bf k}_2, {\bf k}_3) = B_\zeta (k_1, k_2, k_3)$) while the trispectrum $T_\zeta ({\bf k}_1, {\bf k}_2, {\bf k}_3, {\bf k}_4)$ and so on still depend on the whole wavevectors. When relaxing the statistical isotropy condition, we must take into account that $P_\zeta ({\bf k}_1, {\bf k}_2)$ and $B_\zeta ({\bf k}_1, {\bf k}_2, {\bf k}_3)$ do depend on the whole wavevectors and so does $f_{\rm NL} ({\bf k}_1, {\bf k}_2, {\bf k}_3)$ (as well as $\tau_{\rm NL} ({\bf k}_1, {\bf k}_2, {\bf k}_3, {\bf k}_4)$ and $g_{\rm NL} ({\bf k}_1, {\bf k}_2, {\bf k}_3, {\bf k}_4)$); however, by virtue of the reality condition on $\zeta$, the relation $P_\zeta ({\bf k}_1, {\bf k}_2) = P_\zeta (-{\bf k}_1, -{\bf k}_2)$ is met \cite{ValenzuelaToledo:2011fj,dklr}. From now on, by virtue of the statistical homogeneity condition, we will call $P_\zeta ({\bf k}_1, {\bf k}_2)$ just as $P_\zeta ({\bf k}_1)$.

The trispectrum of $\zeta$, $T_\zeta ({\bf k}_1, {\bf k}_2, {\bf k}_3, {\bf k}_4)$, is parameterized, by definition, in terms of the spectrum of $\zeta$, i.e. $P_\zeta ({\bf k})$, and the levels of non-gaussianity $\tau_{\rm NL} ({\bf k}_1, {\bf k}_2, {\bf k}_3, {\bf k}_4)$ and $g_{\rm NL} ({\bf k}_1, {\bf k}_2, {\bf k}_3, {\bf k}_4)$ in the following way \cite{Boubekeur:2005fj,Sasaki:2006kq}:
\begin{eqnarray}
T_\zeta ({\bf k}_1, {\bf k}_2, {\bf k}_3, {\bf k}_4) &=& \tau_{\rm NL} ({\bf k}_1, {\bf k}_2, {\bf k}_3, {\bf k}_4) \left[P_\zeta ({\bf k}_1) P_\zeta ({\bf k}_3) P_\zeta ({\bf k}_1 + {\bf k}_2) + 11 \ {\rm permutations} \right] \nonumber \\
&& + \frac{54}{25} g_{NL} ({\bf k}_1, {\bf k}_2, {\bf k}_3, {\bf k}_4) \left[P_\zeta ({\bf k}_1) P_\zeta ({\bf k}_2) P_\zeta ({\bf k}_3) + 3 \ {\rm permutations} \right] \,. \label{taudef}
\end{eqnarray}
This relation is greatly simplified in the collapsed limit ${\bf k}_1 + {\bf k}_2 \rightarrow 0$ (which indeed implies ${\bf k}_3 + {\bf k}_4 \rightarrow 0$ and, therefore, ${\bf k}_1 \rightarrow - {\bf k}_2$ and ${\bf k}_3 \rightarrow - {\bf k}_4$) and by making use of the reality condition of $\zeta$ and the fact that $P_\zeta ({\bf k}) = \frac{2\pi^2}{k^3} \mathcal{P}_\zeta ({\bf k})$ where $\mathcal{P}_\zeta ({\bf k})$ is almost scale-invariant \cite{wmap,Planck2013XVI} according to observations:
\begin{equation}
\lim_{{\bf k}_1 + {\bf k}_2 \to 0} T_\zeta ({\bf k}_1, {\bf k}_2, {\bf k}_3, {\bf k}_4) = 4 \tau_{\rm NL} ({\bf k}_1, {\bf k}_3) P_\zeta ({\bf k}_1) P_\zeta ({\bf k}_3) P_\zeta ({\bf k}_1 + {\bf k}_2) \,. \label{collapsed}
\end{equation}
In turn, the bispectrum of $\zeta$, $B_\zeta ({\bf k}_1, {\bf k}_2, {\bf k}_3)$, is parameterized, by definition, in terms of the spectrum of $\zeta$, i.e. $P_\zeta ({\bf k})$, and the level of non-gaussianity $f_{\rm NL} ({\bf k}_1, {\bf k}_2, {\bf k}_3)$ in the following way \cite{Komatsu:2001rj,Maldacena:2002vr}:
\begin{equation}
B_\zeta ({\bf k}_1, {\bf k}_2, {\bf k}_3) = \frac{6}{5} f_{\rm NL} ({\bf k}_1, {\bf k}_2, {\bf k}_3) \left[P_\zeta ({\bf k}_1) P_\zeta ({\bf k}_2)  + {\rm cyclic} \ {\rm permutations} \right] \,. \label{fdef}
\end{equation}
The above relation is simplified in the squeezed limit ${\bf k}_1 \rightarrow 0$ (which indeed implies ${\bf k}_2 \rightarrow - {\bf k}_3$) and by making use again of the reality condition of $\zeta$ and the fact that $P_\zeta ({\bf k}) = \frac{2\pi^2}{k^3} \mathcal{P}_\zeta ({\bf k})$:
\begin{equation}
\lim_{{\bf k}_1 \to 0} B_\zeta ({\bf k}_1, {\bf k}_2, {\bf k}_3) = \frac{12}{5} f_{\rm NL} ({\bf k}_2) P_\zeta ({\bf k}_1) P_\zeta ({\bf k}_2) \,. \label{squeezed}
\end{equation}

We will now define an auxiliary field called $\hat{P} ({\bf k})$ as
\begin{equation}
\hat{P} ({\bf k}) = \frac{1}{V_S} \int_{{\bf k}' \in b_S} \frac{d^3 {\bf k}'}{(2\pi)^3} \frac{\zeta({\bf k}') \zeta({\bf k} - {\bf k}')}{P_\zeta ({\bf k}')} \,,
\end{equation}
where $b_S$ is a 
narrow band of wavevectors which are very
near some ${\bf k}_S$ and $V_S$ is its volume:
\begin{equation}
V_S = \int_{{\bf k}' \in b_S} \frac{d^3 {\bf k}'}{(2\pi)^3} \,.
\end{equation}

The next step is to calculate the correlator $\langle \hat{P}^* ({\bf k}_L) \hat{P} ({\bf k}_L) \rangle$: 
\begin{eqnarray}
\langle \hat{P}^* ({\bf k}_L) \hat{P} ({\bf k}_L) \rangle &=& \frac{1}{V_S^2} \int_{{\bf k}_1, {\bf k}_3 \in b_S} \frac{d^3 {\bf k}_1 d^3 {\bf k}_3}{(2\pi)^6} \frac{\langle \zeta(-{\bf k}_1) \zeta(-{\bf k}_L + {\bf k}_1) \zeta({\bf k}_3) \zeta({\bf k}_L - {\bf k}_3) \rangle}{P_\zeta ({\bf k}_1) P_\zeta ({\bf k}_3)} \nonumber \\
&=& \frac{1}{V_S^2} \int_{{\bf k}_1, {\bf k}_3 \in b_S} \frac{d^3 {\bf k}_1 d^3 {\bf k}_3}{(2\pi)^6} \frac{1}{P_\zeta ({\bf k}_1) P_\zeta ({\bf k}_3)} \Big[\langle \zeta(-{\bf k}_1) \zeta(-{\bf k}_L + {\bf k}_1) \zeta({\bf k}_3) \zeta({\bf k}_L - {\bf k}_3) \rangle_c \nonumber \\
&& + \langle \zeta(-{\bf k}_1) \zeta(-{\bf k}_L + {\bf k}_1) \rangle \langle \zeta({\bf k}_3) \zeta({\bf k}_L - {\bf k}_3) \rangle + \langle \zeta(-{\bf k}_1) \zeta({\bf k}_3) \rangle \langle \zeta(-{\bf k}_L + {\bf k}_1)  \zeta({\bf k}_L - {\bf k}_3) \rangle \nonumber \\
&& + \langle \zeta(-{\bf k}_1) \zeta({\bf k}_L - {\bf k}_3) \rangle \langle \zeta(-{\bf k}_L + {\bf k}_1) \zeta({\bf k}_3) \rangle \Big] \,,
\end{eqnarray}
where we have used the Wick's theorem \cite{weinbergbook,wick}. Thus, by employing the statistical homogeneity condition and the reality condition of $\zeta$, $\langle \hat{P}^* ({\bf k}_L) \hat{P} ({\bf k}_L) \rangle$ is finally written as
\begin{eqnarray}
\langle \hat{P}^* ({\bf k}_L) \hat{P} ({\bf k}_L) \rangle &=& \frac{1}{V_S^2} \int_{{\bf k}_1, {\bf k}_3 \in b_S} \frac{d^3 {\bf k}_1 d^3 {\bf k}_3}{(2\pi)^3} \delta^3 ({\bf 0}) \frac{T_\zeta(-{\bf k}_1, -{\bf k}_L + {\bf k}_1, {\bf k}_3, {\bf k}_L - {\bf k}_3)}{P_\zeta ({\bf k}_1) P_\zeta ({\bf k}_3)} \nonumber \\
&& + (2\pi)^6 \left[\delta^3 ({\bf k}_L) \right]^2 + \frac{1}{V_S^2} \int_{{\bf k}_1 \in b_S} d^3 {\bf k}_1 \ \delta^3 ({\bf 0}) \frac{P_\zeta ({\bf k}_1 - {\bf k}_L)}{P_\zeta ({\bf k}_1)} + \frac{1}{V_S^2} \int_{{\bf k}_1 \in b_S} d^3 {\bf k}_1 \ \delta^3 ({\bf 0}) \,.
\end{eqnarray}
The collapsed limit for the trispectrum of $\zeta$ in the above expression is obtained when $-{\bf k}_1 - {\bf k}_L + {\bf k}_1 \rightarrow 0$, i.e. when ${\bf k}_L \rightarrow 0$. Thus, by employing the collapsed limit for the trispectrum of $\zeta$ in Eq. (\ref{collapsed}), and taking the limit ${\bf k}_L \rightarrow 0$ in the above expression, we find
\begin{equation}
\lim_{{\bf k}_L \to 0} \frac{\langle \hat{P}^* ({\bf k}_L) \hat{P} ({\bf k}_L) \rangle}{(2\pi)^3 \delta^3 ({\bf 0})} = 4 P_\zeta ({\bf k}_L) \frac{1}{V_S^2} \int_{{\bf k}_1, {\bf k}_3 \in b_S} \frac{d^3 {\bf k}_1 d^3 {\bf k}_3}{(2\pi)^6} \tau_{\rm NL} ({\bf k}_1, {\bf k}_3) + \frac{2}{V_S} \,. \label{pcorr}
\end{equation}

We will now calculate the correlator $\langle \zeta^*({\bf k}_L) \hat{P} ({\bf k}_L) \rangle$:
\begin{equation}
\langle \zeta^*({\bf k}_L) \hat{P} ({\bf k}_L) \rangle = \frac{1}{V_S} \int_{{\bf k}_2 \in b_S} \frac{d^3 {\bf k}_2}{(2 \pi)^3} \frac{\langle \zeta(-{\bf k}_L) \zeta({\bf k}_2) \zeta({\bf k}_L - {\bf k}_2) \rangle_c}{P_\zeta ({\bf k}_2)} \,.
\end{equation}
The squeezed limit for the bispectrum of $\zeta$ in the above expression is obtained when ${\bf k}_L \rightarrow 0$. Thus, by employing the squeezed limit for the bispectrum of $\zeta$ in Eq. (\ref{squeezed}), and taking the limit ${\bf k}_L \rightarrow 0$ in the above expression, we find
\begin{equation}
\lim_{{\bf k}_L \to 0} \frac{\langle \zeta^* ({\bf k}_L) \hat{P} ({\bf k}_L) \rangle}{(2\pi)^3 \delta^3 ({\bf 0})} = \frac{12}{5} P_\zeta ({\bf k}_L) \frac{1}{V_S} \int_{{\bf k}_2 \in b_S} \frac{d^3 {\bf k}_2}{(2\pi)^3} f_{\rm NL} ({\bf k}_2) \,. \label{crosscorr}
\end{equation}

Finally, let's construct the covariance matrix
\begin{equation}
\frac{1}{(2\pi)^3 \delta^3 ({\bf 0})}
\begin{pmatrix}
\langle \zeta^* ({\bf k}_L) \zeta({\bf k}_L) \rangle & \langle \zeta^* ({\bf k}_L) \hat{P} ({\bf k}_L) \rangle \\
\langle \zeta ({\bf k}_L) \hat{P}^* ({\bf k}_L) \rangle & \langle \hat{P}^* ({\bf k}_L) \hat{P} ({\bf k}_L) \rangle
\end{pmatrix} \,,
\end{equation}
whose determinant is positive;  indeed, such a condition, in the ${\bf k}_L \rightarrow 0$ limit, and by introducing Eqs. (\ref{pcorr}) and (\ref{crosscorr}), reduces to
\begin{equation}
\frac{1}{V_S^2} \int_{{\bf k}_1, {\bf k}_3 \in b_S} \frac{d^3 {\bf k}_1 d^3 {\bf k}_3}{(2\pi)^6} \tau_{\rm NL} ({\bf k}_1, {\bf k}_3) \geqslant \left[ \frac{6}{5} \frac{1}{V_S} \int_{{\bf k}_2 \in b_s} \frac{d^3 {\bf k}_2}{(2\pi)^3}  f_{\rm NL} ({\bf k}_2) \right]^2 - \frac{1}{2 P_\zeta ({\bf k}_L) V_S} \,,  
\end{equation}
where the last term in the right hand side of the equation goes to zero in the ${\bf k}_L \rightarrow 0$ limit.  Thus, we arrive to the actual expression obtained by Smith, LoVerde, and Zaldarriaga \cite{cr2}, before assuming scale invariance, but this time relaxing the statistical isotropy condition:
\begin{equation}
\boxed{\int_{{\bf k}_1, {\bf k}_3 \in b_s} \frac{d^3 {\bf k}_1 d^3 {\bf k}_3}{(2\pi)^6} \tau_{\rm NL} ({\bf k}_1, {\bf k}_3) \geqslant \left[ \frac{6}{5} \int_{{\bf k}_2 \in b_s} \frac{d^3 {\bf k}_2}{(2\pi)^3}  f_{\rm NL} ({\bf k}_2) \right]^2} \,. \label{slzconsistency}
\end{equation}
When statistical isotropy is required, the argument of $f_{\rm NL}$ in the above inequality must be replaced so that it becomes the wavenumber $k_2$ instead of the wavevector ${\bf k}_2$. We will call Eq. (\ref{slzconsistency}) the second variety of the SY consistency relation.

It is not our intention to reproduce in this paper the proof by Assassi, Baumann, and Green \cite{abg}, but we can say that, relaxing the statistical isotropy condition, and before assuming scale invariance, their proof leads actually to the consistency relation
\begin{equation}
\boxed{\int \frac{d^3{\bf k}_1 d^3{\bf k}_3}{(2\pi)^6} \tau_{\rm NL} ({\bf k}_1, {\bf k}_3) P_\zeta ({\bf k}_1) P_\zeta ({\bf k}_3)  \geqslant \left[ \frac{6}{5} \int \frac{d^3 {\bf k}_2}{(2\pi)^3}  f_{\rm NL} ({\bf k}_2) P_\zeta ({\bf k}_2) \right]^2} \,,  \label{abgconsistency}
\end{equation}
where the integrals are performed over the entire momentum space. When statistical isotropy is required, the arguments of $P_\zeta$ and $f_{\rm NL}$ in the above inequality must be replaced so that they become the respective wavenumbers $k_n$ instead of the respective wavevectors ${\bf k}_n$, fixing in this way a small mistake in Eq. (2.13) of \cite{abg}. We will call Eq. (\ref{abgconsistency}) the third variety of the SY consistency relation.

\section{The scale invariance argument:  the fourth and fifth varieties of the SY consistency relation}

The previous section showed what the actual consistency relations are, between $\tau_{\rm NL}$ and $f_{\rm NL}$, before assuming scale invariance.  If scale invariance is guaranteed, as argued by Kehagias and Riotto \cite{kr} in the case that the inflationary dynamics is driven by Lorentz-invariant scalar fields only, the second and the third varieties of the SY consistency relation, Eqs. (\ref{slzconsistency}) and (\ref{abgconsistency}), reduce to the first variety in Eq. (\ref{sye}).  However, why should it be like that in the general case?  Could it be that, although the second and the third varieties are satisfied, the first variety is violated for those cases where there is no scale invariance?, i.e. is it possible to have the following relation?:
\begin{equation}
\tau_{\rm NL} ({\bf k}_1, {\bf k}_3) < \left(\frac{6}{5}\right)^2 f_{\rm NL} ({\bf k}_1) f_{\rm NL} ({\bf k}_3) \,. \label{SYvio}
\end{equation}
It is worth noticing that the previous expression corresponds to a violation of what we will call the fourth variety of the SY consistency relation:
\begin{equation}
\boxed{\tau_{\rm NL} ({\bf k}_1, {\bf k}_3) \geqslant \left(\frac{6}{5}\right)^2 f_{\rm NL} ({\bf k}_1) f_{\rm NL} ({\bf k}_3)} \,, \label{mpsy}
\end{equation}
which is just a direct generalization of the first variety, Eq. (\ref{sye}), when there is no scale-invariance and whose form is easily inspired from the second and third varieties, Eqs. (\ref{slzconsistency}) and (\ref{abgconsistency}).
Well, it is clear that, if the proof by Kehagias and Riotto \cite{kr} is valid to all orders, the violation of the fourth variety of the SY consistency relation would imply the presence of new non-trivial degrees of freedom (see for example Refs. \cite{kr,Choudhury:2012kw});  however, the opposite is not necessarily true, and that is the motivation of our mission in the next section: to show that, when involving vector fields, there exist models where the inequality in Eq. (\ref{mpsy}) is indeed strongly  violated for some wavevector configurations. Meanwhile, what we can say is that the fourth variety of the SY consistency relation is satisfied when the arguments of $\tau_{\rm NL}$ and $f_{\rm NL}$ are all the same even if the scale invariance is not guaranteed:
\begin{equation}
\boxed{\tau_{\rm NL} ({\bf k}_1, {\bf k}_1) \geqslant \left(\frac{6}{5} f_{\rm NL} ({\bf k}_1) \right)^2} \,; \label{5vsy}
\end{equation}
this can be proven by observing that, in Eq. (\ref{slzconsistency}), $b_S$ can be made as small as we want, getting rid of the integrals, but making the arguments of $\tau_{\rm NL}$ and $f_{\rm NL}$ all the same. We will call Eq. (\ref{5vsy}) the fifth variety of the SY consistency relation.  The discussions in Sections \ref{first} and \ref{second} of this paper show that statistical homogeneity is actually the only non-general condition for the validity of the fifth variety of the SY consistency relation when $\tau_{\rm NL} ({\bf k}_1, {\bf k}_2, {\bf k}_3, {\bf k}_4)$ and $f_{\rm NL} ({\bf k}_1, {\bf k}_2, {\bf k}_3)$ are evaluated in the collapsed and squeezed limits respectively; an observed violation of such a variety would lead to the conclusion that statistical homogeneity does not hold and that, therefore, theory and observation cannot be compared \cite{ValenzuelaToledo:2011fj} leading to a strong shaking of the foundations of cosmology as a science \cite{yrodrig}.

\section{A class of inflationary models involving vector fields: violating the fourth variety of the SY consistency relation} \label{fourth}
In this section we show a worked example in which we get a violation of the fourth variety of the SY consistency relation in Eq. (\ref{mpsy}) at least at tree level. 
We begin by describing a class of inflationary models which includes several popular models of inflation in the presence of vector fields;  this will be done using a generic parametrization.

\subsection{A class of inflationary models involving vector fields in view of the $\delta N$ formalism}
The class of models that we shall consider in this section can be parametrized using its first three correlation functions. We use the $\delta N$ formalism \cite{ValenzuelaToledo:2011fj,Lyth:2005fi} to do our calculations at tree level for all the correlators and we will assume isotropic expansion. In the following, we shall use the notation $A_{\bar{I}}=(\phi_{I}, {\bf A}^{(a)}_{i})$ for the fields, where the indices $\bar{I}, \bar{J}, \dots $ label both scalar and vector fields; for scalar fields we use capital latin indices $ I, J, \dots$ and for vector fields we use lower case latin indices $i, j,\dots$ for vector components and another set of supra indices $a, b, \dots$ to label the vector field \cite{ValenzuelaToledo:2011fj}. In the following example we consider models involving only one vector field so we can drop the supra index $a,b, \dots$  For the derivatives of the amount of expansion $N$, we use:
\be 
N_{\bar{I}} = \frac{\partial N}{\partial A_{\bar{I}}}, \; N_{\bar{I} \bar{J}} = \frac{\partial^{2} N}{\partial A_{\bar{I}}\partial A_{\bar{J}}}, \; \mbox{etc.}
\ee
For simplicity, we shall consider models in which there is only one vector field so there is only one preferred direction given by the unit vector ${\bf \hat{{\bf n}}}_{i} = N_{i}/(N_{k}N_{k})^{1/2}$;  however, the generalization for several scalar and vector fields should be straightforward (for more details, see Ref. \cite{BeltranAlmeida:2011db}). We will also impose the following conditions over the derivatives of $N$ \footnote{These conditions are rather typical when the background space undergoes isotropic expansion \cite{dklr}. One could argue, in principle, that these conditions still hold even when the background undergoes anisotropic expansion given that the potentials of the fields obey suitable slow-roll conditions. }:
\be\label{conditions} 
N_{\bar{I}} \propto A_{\bar{I}}, \; N_{\bar{I}\bar{J}}  \propto \delta_{\bar{I}\bar{J}}. 
\ee
The second condition impose that the interactions between scalar and vector fields are null or at least are negligible compared to the scalar-scalar and vector-vector interactions. Some of the most popular inflationary models in the presence of vector fields such as vector curvaton \cite{vi2}-\cite{vi3}, vector inflation \cite{Golovnev:2008cf}, and the hybrid inflation with coupled vector and scalar ``waterfall field'' where $\zeta$ is generated at the end of inflation \cite{vi} are indeed parametrized as we present here. These cases certainly obey the conditions stated in Eq. (\ref{conditions}).   With the notation and conditions described above, the power spectrum of $\zeta$ is parametrized as follows \cite{Ackerman:2007nb}:
\be \label{ps} { P}_{\zeta}({\bf k})= { P}_{\zeta}
^{{\rm iso}}(k)\left(1+g_\zeta(\hat{{\bf k}}\cdot \hat{{\bf n}})^2 \right) \,,
\ee
 where 
\begin{equation}
{ P}_{\zeta}^{{\rm iso}}(k)=N_{a}N_{b}P_{ab}(k) \,,
\end{equation}
is the isotropic part of the power spectrum with $P_{ab}$ being the power spectra of the field perturbations. The dimensionless power spectrum is related to the spectra of scalar and vector perturbations through $2\pi^2 {\cal P}_{ab}(k) = k^{(3-n_{ab})} P_{ab}(k)$. For simplicity we assume that all the perturbations have approximately the same spectral index so that we take $n_{ab} \approx n_{s}$. In any case, the results we exhibit here are not affected by the inclusion of this index and we could neglect it. 
The vector $\hat{\bf k}$ is the unit vector along the wavevector $\bf{k}$, the parameter $g_{\zeta}$ quantifies the level of statistical anisotropy in $\zeta$, and the vector $\hat{\bf n}$ specifies the direction of this anisotropy. In general, the  statistical anisotropy parameter $g_\zeta$ should acquire some level of scale dependence; however, in order to simplify our analysis, we consider here a scale-invariant anisotropy parameter.
 
The bispectrum, expanded in $g_\zeta$, is parametrized as follows\footnote{See Ref. \cite{Bartolo:2011ee} for a detailed discussion of the type of models that can be parametrized in this way. In that reference, the authors consider a scale-dependent  anisotropy parameter $g_{\zeta}(k)$ and even present a more general parametrization which involves non-Abelian gauge fields. More recently, in Ref. \cite{Shiraishi:2013vja} the authors investigate the phenomenological consequences of a more general parametrization based on a Legendre polynomials expansion.}:
\ba \nonumber
 B_{\zeta}({\bf k}_1 ,\,{\bf k}_2 ,\, {\bf k}_3 ) & \equiv &  (1+\xi_1)B_{\zeta}^{{\rm iso}}({ k}_1 ,\,{ k}_2 ,\, { k}_3 ) \\ \nonumber 
&=& \,\left[ 1+\bar{g}_{\zeta}\, f \,\frac{\sum\limits_{l<m}\, 
(k_l k_m)^{-(3-n_{s})}\left[ (\hat{{\bf n}}\cdot \hat{{\bf k}}_{(l)} )^{2}  + (\hat{{\bf n}}\cdot \hat{{\bf k}}_{(m)} )^{2} \right]}{ \sum\limits_{l<m} (k_l k_m)^{-(3-n_{s})}} \, \right. \\ 
 &+&   \left. \bar{g}_{\zeta}^2\, f\, \frac{  \sum\limits_{l<m}
(k_l k_m)^{-(3-n_{s})}  (\hat{{\bf n}}\cdot \hat{{\bf k}}_{(l)}) (\hat{{\bf n}}\cdot \hat{{\bf k}}_{(m)}) (\hat{{\bf k}}_{(l)} \cdot \hat{{\bf k}}_{(m)})  }{\sum\limits_{l<m} (k_l k_m)^{-(3-n_{s})} } \right] B_{\zeta}^{{\rm iso}}({ k}_1 ,\,{ k}_2 ,\, { k}_3 ) \,, \label{bs}
\ea 
where $\hat{{\bf k}}_{(l)i} $, with $i=1,2,3$, denote the components of the vectors ${\hat{{\bf k}}}_1 ,\,{\hat{{\bf k}}}_2 ,\, {\hat{{\bf k}}}_3$,
\begin{equation}
B_{\zeta}^{{\rm iso}}({ k}_1 ,\,{ k}_2 ,\, { k}_3 ) = N_a N_b N_{cd}{\cal P}_{ac}{\cal P}_{bd} \,  \sum_{l<m} (k_l k_m)^{-(3-n_{s})} \,,
\end{equation}  
$\bar{g}_{\zeta} =  \left(\frac{{N}_a {N}_b {\cal P}_{ab}   }{ {N}_i {N}_i {\cal P}_{+}  }\right)g_{\zeta}$,  and $f =  \left(\frac{{N}_i {N}_i {\cal P}_{+}^{2}   }{ {N}_a {N}_b {\cal P}_{ac} {\cal P}_{bc}  }\right)$.  In the above expressions, ${\cal P}_+$ is the parity-conserving power spectrum defined as ${\cal P}_+ = \frac{{\cal P}_R + {\cal P}_L}{2}$ where ${\cal P}_{R,L}$ are the right and left vector field spectra in a circular polarization basis (for details, see Refs. \cite{ValenzuelaToledo:2011fj,dklr}).
 
Finally, the trispectrum, expanded in  $g_\zeta$, is parametrized as follows:
\ba\label{ts} \nonumber
&& T_{\zeta}^{\tau_{\rm NL}}({\bf k}_1 ,\,{\bf k}_2 ,\, {\bf k}_3 , \, {\bf k}_4 )  \equiv   (1+\xi_2)T_{\zeta}^{{\rm iso}}({ k}_1 ,\,{ k}_2 ,\, { k}_3 ,\, { k}_4) \\ \nonumber 
&& = \,\left[ 1+\bar{g}_{\zeta} \,h \,\frac{  \sum\limits_{l<m, s\neq m, l} (k_l k_m k_{ls})^{-(3-n_{s})} \left[  (\hat{{\bf n}}\cdot \hat{{\bf k}}_{(l)} )^{2}  + (\hat{{\bf n}}\cdot \hat{{\bf k}}_{(m)} )^{2} +    (\hat{{\bf n}}\cdot \hat{{\bf k}}_{(ls)} )^{2} \right]}{  \sum\limits_{l<m, s\neq m, l} (k_l k_m k_{ls})^{-(3-n_{s})}} \, \right. \\ \nonumber 
&&  + \left. \bar{g}_{\zeta}^2\, h \, \frac{ \sum\limits_{l<m, s\neq m, l} (k_l k_m k_{ls})^{-(3-n_{s})} \left[ (\hat{{\bf n}}\cdot \hat{{\bf k}}_{(l)}) (\hat{{\bf n}}\cdot \hat{{\bf k}}_{(m)}) (\hat{{\bf k}}_{(l)} \cdot \hat{{\bf k}}_{(m)})  \right]}{  \sum\limits_{l<m, s\neq m, l} (k_l k_m k_{ls})^{-(3-n_{s})} }   \right. \\ \nonumber
&&  + \left. \bar{g}_{\zeta}^2\, h \,  \frac{ \sum\limits_{l<m, s\neq m, l} (k_l k_m k_{ls})^{-(3-n_{s})}   (\hat{{\bf n}}\cdot \hat{{\bf k}}_{(ls)})\left[    (\hat{{\bf n}}\cdot \hat{{\bf k}}_{(m)}) (\hat{{\bf k}}_{(m)}\cdot \hat{{\bf k}}_{(ls)}) +  (\hat{{\bf n}}\cdot \hat{{\bf k}}_{(l)}) (\hat{{\bf k}}_{(l)}\cdot \hat{{\bf k}}_{(ls)})  \right]}{  \sum\limits_{l<m, s\neq m, l} (k_l k_m k_{ls})^{-(3-n_{s})} }   \right. \\ 
&&  +  \left. \bar{g}_{\zeta}^3 \, h\,  \frac{ \sum\limits_{l<m, s\neq m, l} (k_l k_m k_{ls})^{-(3-n_{s})}   (\hat{{\bf n}}\cdot \hat{{\bf k}}_{(l)})  (\hat{{\bf n}}\cdot \hat{{\bf k}}_{(m)}) (\hat{{\bf k}}_{(l)}\cdot \hat{{\bf k}}_{(ls)})(\hat{{\bf k}}_{(m)}\cdot \hat{{\bf k}}_{(ls)}) }{  \sum\limits_{l<m, s\neq m, l} (k_l k_m k_{ls})^{-(3-n_{s})} } \right] T_{\zeta}^{{\rm iso}}({ k}_1 ,\,{ k}_2 ,\, { k}_3, \, { k}_4  ) \,,
\ea
where we use the notation $k_{mn} = |\bf{k}_m + \bf{k}_n|$, $h= \left(\frac{{N}_i {N}_i {\cal P}_{+}^{3}   }{ {N}_a {N}_b {\cal P}_{ad} {\cal P}_{bf}{\cal P}_{df}  }\right)$  and
\begin{equation}
T_{\zeta}^{{\rm iso}}({ k}_1 ,\,{ k}_2 ,\, { k}_3,\, { k}_4 )\,=\, N_a N_b N_{cd} N_{ef}\, {\cal P}_{ac} {\cal P}_{be} {\cal P}_{df}\,  \sum_{l<m,  s\neq m, l} (k_l k_m k_{ls})^{-(3-n_{s})} \,.
\end{equation}
\subsection{The non-gaussianity parameters}
\begin{figure}[h!]
        \centering
        \begin{subfigure}[b]{0.51\textwidth}
                \centering
                \includegraphics[width=\textwidth]{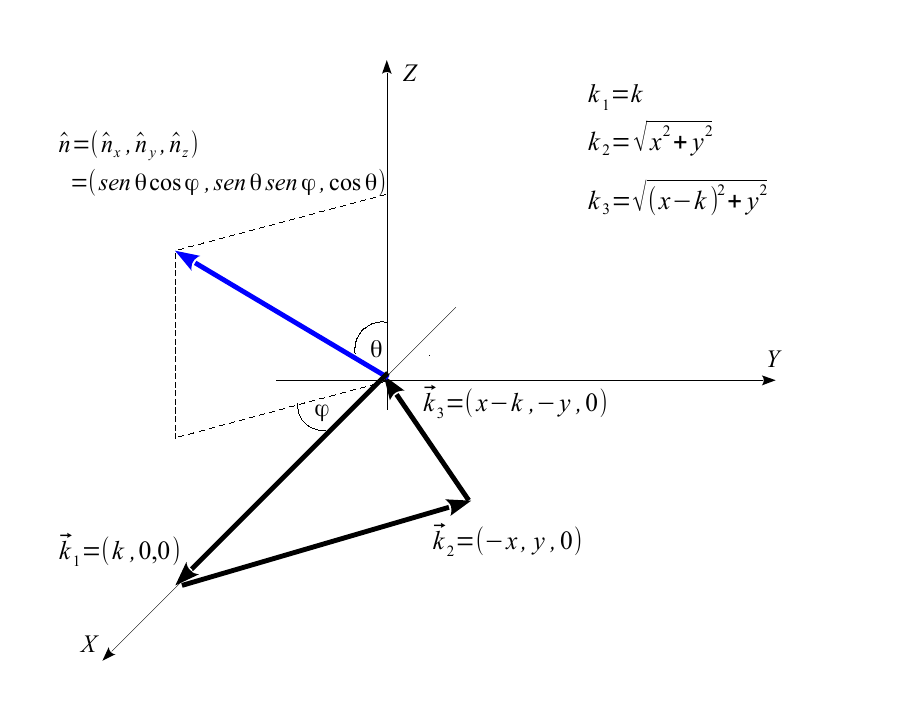}
                \caption{Parametrization of the bispectrum of $\zeta$ and $f_{\rm NL}$.}
                \label{BS-par}
        \end{subfigure}%
        ~ 
        \begin{subfigure}[b]{0.51\textwidth}
                \centering
                \includegraphics[width=\textwidth]{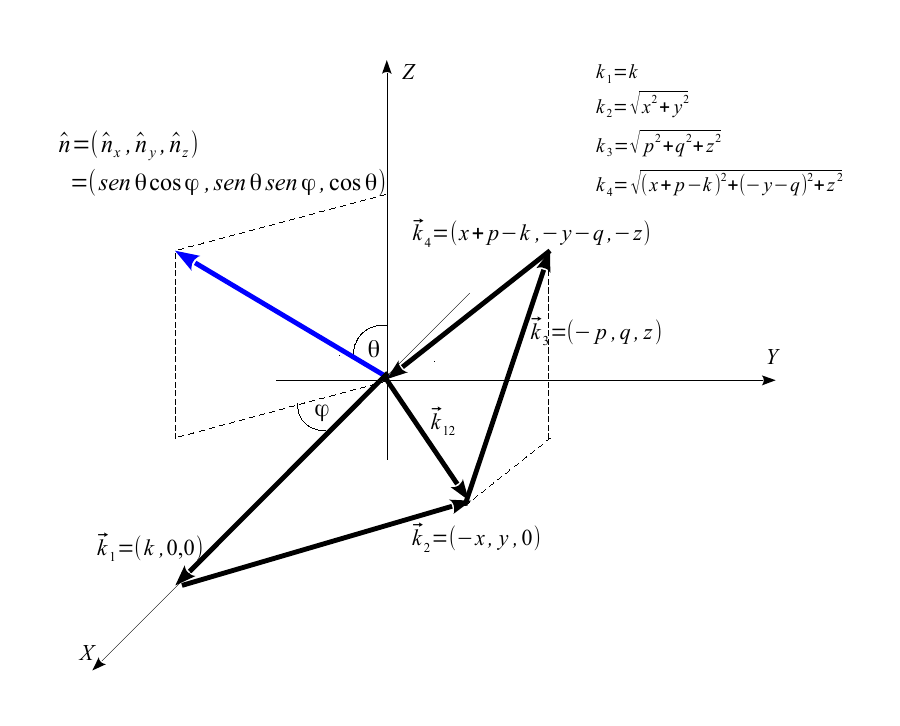}
                \caption{Parametrization of the trispectrum of $\zeta$ and $\tau_{\rm NL}$.}
                \label{TS-par}
        \end{subfigure}
\caption{The figures (a) and (b) represent the parameters employed to calculate the correlation functions of $\zeta$ and the non-gaussianity parameters. Aside of the wavevectors, when statistical anisotropy is taken into account, one needs a unitary direction of a preferred direction. Here, this direction is represented by the vector ${\bf \hat{{\bf n}}}$ which is parametrized using spherical coordinates $(\theta, \phi)$.   
}
\label{parametrization}
\end{figure}

The non-Gaussianity parameters $f_{\rm NL}$ and $\tau_{\rm NL}$ defined through Eqs. (\ref{fdef}) and (\ref{taudef})
can be obtained directly from Eqs. (\ref{ps}), (\ref{bs}), and (\ref{ts}). Since our intention is to describe cases in which the fourth variety of the SY consistency relation in Eq. (\ref{mpsy}) is violated, a simple minded example is enough for our purposes; then, we consider a single vector field so that $\bar{g}_{\zeta} = g_{\zeta}$ and $f=h=1$. 
For our evaluations, we use the parametrization described in Fig. \ref{parametrization}. As shown in the figure, the wavevectors ${\bf{k}}_{i}$ are restricted to form a warped polygon due to the statistical homogeneity and the vector $\hat{{\bf n}}$ is described by its polar and azimuthal angles $\theta$ and $\phi$ respectively. In sum, the bispectrum of $\zeta$ and the level of non-gaussianity $f_{\rm NL}$ are functions of the lengths of the triangle formed by the wavevectors, the angles that parametrize the orientation of $\hat{{\bf n}}$, and the level of statistical anisotropy $g_{\zeta}$; this is: $f_{\rm NL } = f_{\rm NL }(k_{1}, k_{2}, k_{3}, \theta, \phi, g_{\zeta})$. Analogously, the trispectrum of $\zeta$ and the level of non-gaussianity $\tau_{\rm NL}$ have the following functional dependence:  $\tau_{\rm NL } = \tau_{\rm NL }(k_{1}, k_{2}, k_{3}, k_{4}, k_{12}, k_{14}, \theta, \phi, g_{\zeta})$. Thus, following Fig. \ref{parametrization}, $f_{\rm NL } = f_{\rm NL }(k, x, y, \theta, \phi, g_{\zeta})$ and $\tau_{\rm NL } = \tau_{\rm NL }(k, x, y, p, q, z,  \theta, \phi, g_{\zeta})$.

\subsection{Results}
\begin{figure}[h!]
        \centering
        \begin{subfigure}[l]{0.5\textwidth}
             \centering
                \includegraphics[width=\textwidth]{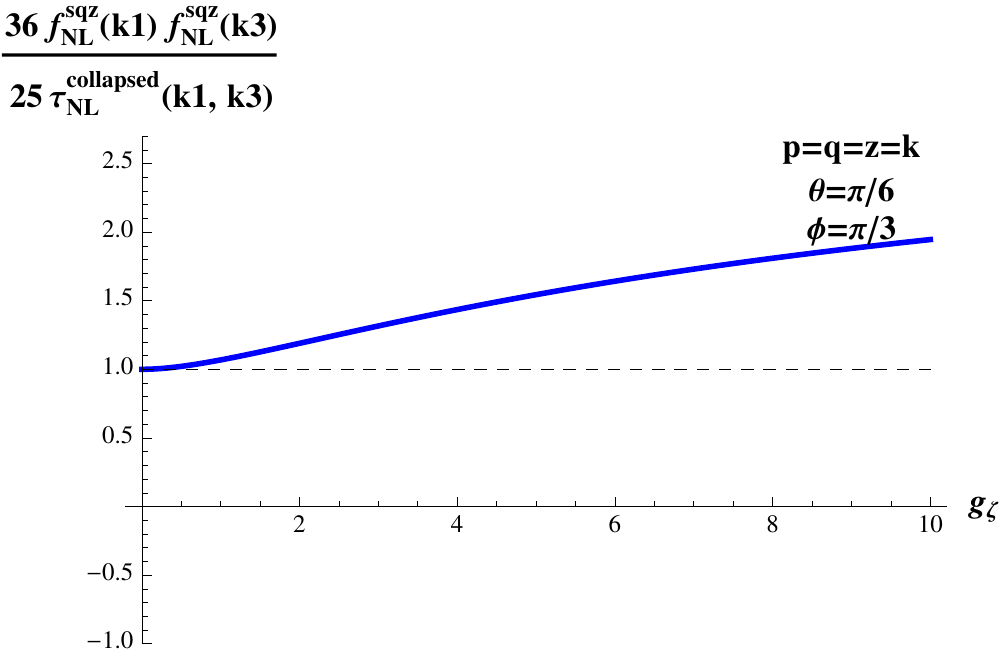}
                \caption{There exist configurations in which the fourth variety of the SY consistency relation (Eq. \ref{mpsy}) is violated. Here, as an example, the configuration $p=q=z=x=k$, $\theta= \pi/6$, and $\phi= \pi/3 $ is showed.}
                \label{k1-k3}
        \end{subfigure}%
        ~ 
        \begin{subfigure}[r]{0.5\textwidth}
               \centering
                \includegraphics[width=\textwidth]{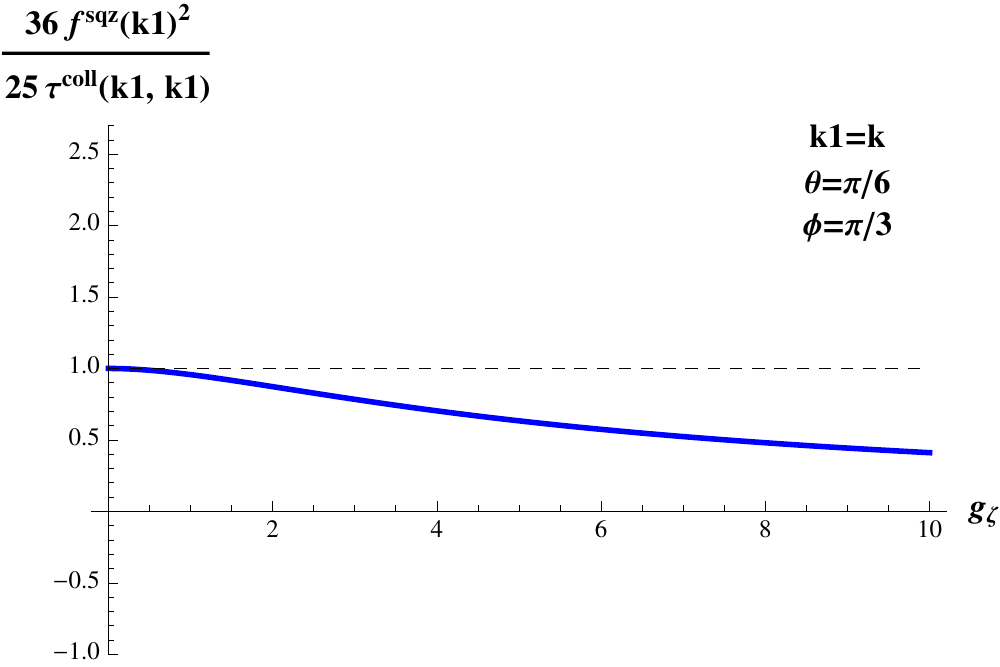}\caption{The fifth variety of the SY consistency relation (Eq. \ref{5vsy}) is preserved in any wavevector configuration. As an example, this figure shows the evaluation of Eq. (\ref{5vsy}) in the configuration $k_{1}=k$, $\theta= \pi/6$, and $\phi= \pi/3$. 
           }
                \label{k1-k1}
        \end{subfigure}
\caption{The fourth and fifth varieties of the SY consistency relation under test.}\label{results}
\end{figure}

In order to test the fourth and fifth varieties of the SY consistency relation, Eqs. (\ref{mpsy}) and (\ref{5vsy}), we have to take the squeezed limit of $f_{\rm NL}$, Eq. (\ref{squeezed}),  and the collapsed limit of $\tau_{\rm NL}$, Eq. (\ref{collapsed}). The squeezed limit of $f_{\rm NL}$ has to be evaluated both when $\bf{k}_1 \to 0$ and when $\bf{k}_3 \to 0$. Then, looking at Fig. \ref{parametrization}(a) and using the parametrization $f_{\rm NL }(k, x, y, \theta, \phi, g_{\zeta})$ we take the first limit as:
\be
f^{\rm sqz}_{\rm NL}({\bf k_1}) \equiv \lim_{{\bf k}_3 \to 0} f_{\rm NL}({\bf k_1}, {{\bf k}_2}, {{\bf k}_3}) =  f_{\rm NL}({{\bf k}_1}, -{{\bf k}_1}, 0) = f_{\rm NL}(k, k, 0, \theta, \phi, g_{\zeta}) \,. 
\ee
For the second limit, we look at Fig. \ref{parametrization}(b) and use the parametrization $f_{\rm NL }(k, x, y, p, q, z, \theta, \phi, g_{\zeta})$:
\be
f^{\rm sqz}_{\rm NL}({{\bf k}_3}) \equiv \lim_{{\bf k}_{12} \to 0} f_{\rm NL}({{\bf k}_{12}}, {{\bf k}_3}, {{\bf k}_4}) =  f_{\rm NL}(0, {{\bf k}_3}, -{{\bf k}_3}) = f_{\rm NL}(k, k, 0, p, q, z, \theta, \phi, g_{\zeta}) \,.
\ee
As a result, both limits are independent of the scale $k$. On the other hand, in the collapsed limit of $\tau_{\rm NL}$, when ${\bf k}_{12}\to 0$,  we obtain: 
\ba
\tau^{\rm coll}_{\rm NL}({{\bf k}_1}, {{\bf k}_3}) \equiv \lim_{{\bf k}_{12} \to 0} \tau_{\rm NL}({{\bf k}_1}, {{\bf k}_2}, {{\bf k}_3}, {{\bf k}_4}) &=&   \tau_{\rm NL}({{\bf k}_1}, -{{\bf k}_1}, {{\bf k}_3}, -{{\bf k}_3}) = \tau_{\rm NL}(k, k, 0, p, q, z,  \theta, \phi, g_{\zeta}) \,,
\ea
which is also $k$-independent. Finally, the collapsed limit evaluated at ${\bf k}_1$ is obtained from the previous expression in the following way:
\ba
\tau^{\rm coll}_{\rm NL}({{\bf k}_1}, {{\bf k}_1}) \equiv \lim_{{\bf k}_{3} \to {\bf k}_{1}}  \tau_{\rm NL}({{\bf k}_1}, -{{\bf k}_1}, {{\bf k}_1}, -{{\bf k}_1})  &=&   \tau_{\rm NL}(k, k, 0, k, 0, 0,  \theta, \phi, g_{\zeta}),
\ea
which is again $k$-independent. The results of $(\frac{6 }{5})^{2}(f^{\rm sqz}_{\rm NL}({\bf k}_{1} ) f^{\rm sqz}_{\rm NL}({\bf k}_{3} ))/\tau^{\rm coll}_{\rm NL}({\bf k}_{1}, {\bf k}_{3}) $ and  $(\frac{6 }{5}f^{\rm sqz}_{\rm NL}({\bf k}_{1} ))^{2}/\tau^{\rm coll}_{\rm NL}({\bf k}_{1}, {\bf k}_{1}) $  evaluated in a particular configuration  are shown in Fig. \ref{results}.  
The results shown in the figure correspond to the case in which all the scales are equal, this is: $p=q=z=x=k$, and the unit vector along the preferred direction is characterized by $\theta= \pi/6$ and $\phi= \pi/3 $. Both quotients in this configuration are independent on the scale $k$ but exhibit a non trivial dependence on the orientation of the unit vector $\hat{{\bf n}}$ and the level of statistical anisotropy $g_{\zeta}$.  In Fig. \ref{results}(a), we have an explicit example of the violation of the fourth variety of the SY consistency relation, Eq. (\ref{mpsy}), while in Fig.  \ref{results}(b), we have an example of the preservation of the fifth variety of the SY consistency relation, Eq. (\ref{5vsy}). The case represented in Fig.  \ref{results}(a) is the relevant result for our purposes because this is the only proof that we need in order to show that the inequality  (\ref{mpsy}) is not a generic feature of models with statistical anisotropy. There are many other (actually infinite) configurations in which violations of the fourth variety, Eq. (\ref{mpsy}), are obtained. The result shown in Fig. \ref{results}(b), on the other hand, is only a particular case of the fifth variety, Eq. (\ref{5vsy}), but we can see that this relation is respected in any configuration of the wavevectors and the vector $\hat{{\bf n}}$ and for any value of the level of statistical anisotropy $g_\zeta$.  

It is worth noticing that we do not consider just the values of $g_{\zeta}$ in agreement with current observational bounds ($g_{\zeta} = 0.290 \pm 0.031$) \cite{gr}  since it is not our intention to restrict the analysis to realistic models of statistical anisotropy, being that the reason why in our evaluations we indeed consider larger values of $g_{\zeta}$, from $0$ to $10$; this is done because it helps us to realize that no matter how large the statistical anisotropy in $\zeta$ can be, the fifth variety of the SY consistency relation, Eq. (\ref{5vsy}), is always respected, which means that this relation is independent on the assumption of statistical isotropy in the primordial curvature perturbation. On the other hand, it is also worth noticing that, even for values of $g_{\zeta}$ in agreement with current observations, we can obtain violations of the fourth variety of the SY consistency relation, Eq. (\ref{mpsy}), within percent level size.

\section{The primordial curvature perturbation generated by scalar and vector fields: the sixth variety of the SY consistency relation}
We briefly present in this section the latest variety of the SY consistency relation which was formally demonstrated in a quite recent paper by the authors of this work \cite{BeltranAlmeida:2011db}:
\begin{equation}
\boxed{\tau^{\rm iso}_{\rm NL} \geqslant \left( \frac{6}{5} f^{\rm iso}_{\rm NL} \right)^2} \,.  \label{syvf}
\end{equation}
In the previous expression, $\tau^{\rm iso}_{\rm NL}$ and $f^{\rm iso}_{\rm NL}$ correspond to the isotropic pieces (i.e., they do not depend at all on the wavevectors) of the levels of non-gaussianity in models where the primordial curvature perturbation is generated by scalar and vector fields.  One could think that $\tau^{\rm iso}_{\rm NL}$ and $f^{\rm iso}_{\rm NL}$, being isotropic pieces, have contributions only from the scalar fields, but, as demonstrated in \cite{BeltranAlmeida:2011db}, they receive contributions from the vector fields as well.

As done in Section 1, we will enumerate the conditions under which the sixth variety of the SY consistency relation is satisfied:
\begin{itemize}
\item {\it Condition 1}: The calculation of $f^{\rm iso}_{\rm NL}$ and $\tau^{\rm iso}_{\rm NL}$ is performed at tree level in the diagrammatic approach of the $\delta N$ formalism \cite{bksw,ValenzuelaToledo:2011fj}.
\item {\it Condition 2}: The expansion is assumed to be isotropic so that the field perturbations (actually the scalar perturbations that multiply the respective polarization vectors (if any) in a polarization mode expansion) are statistically isotropic.
\item {\it Condition 3}: The inflationary dynamics is driven by any number of slowly-rolling vector and scalar fields.
\item {\it Condition 4}: The fields involved are gaussian.
\item {\it Condition 5}: The field perturbations (with the proviso expressed in {\it Condition 2}) are scale-invariant.
\end{itemize}
As discussed previously, $f_{\rm NL}$ and $\tau_{\rm NL}$ are not directly comparable in the general case because they are functions of the wavevectors; however, the expression in Eq. (\ref{syvf}) is valid since, under the conditions just stated above, $f^{\rm iso}_{\rm NL}$ and $\tau^{\rm iso}_{\rm NL}$ are scale-invariant \cite{dklr}.

\section{Conclusions: The SY consistency relation(s), not just one but six}
When first introduced back in 2008 \cite{Suyama:2007bg}, the SY consistency relation was considered as a useful tool to discriminate among different classes of models for the generation of the primordial curvature perturbation;  however, the quite strong conditions under which this relation was proven led to several authors to wonder whether such conditions could be relaxed in the search of a more accurate and clean way to employ this consistency relation as a discriminator tool.  Since that time, six different varieties of the SY consistency relation have been implicitly introduced in the literature, one of them indeed introduced in the present paper.  In this work, we have collected the six varieties and showed them explicitly as well as the conditions under which they are valid.  The first and the sixth varieties, Eqs. (\ref{sye}) and (\ref{syvf}), are quite similar in the sense that the non-gaussianity parameters $f_{\rm NL}$ and $\tau_{\rm NL}$ are scale-invariant;  they are different in the sense that the first variety involves the whole $f_{\rm NL}$ and $\tau_{\rm NL}$ while the sixth variety involves only their isotropic pieces.  The second and third varieties of the SY consistency relation, Eqs. (\ref{slzconsistency}) and (\ref{abgconsistency}), are very similar since they involve an integration over the momentum space but are different because the integration in the second variety is just in a narrow band of wavevectors whereas in the third variety the integration is over the entire momentum space while introducing the power spectrum of the primordial curvature perturbation as part of the integrand;  anyway, because of the integrations, it looks quite difficult to employ these varieties of the SY consistency relation in order to discriminate among different classes of models.  What is interesting however of these varieties is that they are valid even if there is neither statistical isotropy nor scale-invariance:  that is why the fourth variety in Eq. (\ref{mpsy}) is so interesting, being just the straightforward generalization of the first variety and whose form is easily inspired from the second and third varieties;  it is for sure valid for models involving only slowly-rolling Lorentz-invariant scalar fields, as long as the proof by Kehagias and Riotto \cite{kr} turns out to be valid at all orders in perturbation theory, but for sure fails in some model that involves non-trivial degrees of freedom;  indeed, we have shown in Section \ref{fourth} the first counterexample: the fourth variety is strongly violated in a quite generic class of models that involves vector fields and that include the vector curvaton model proposed by Dimopoulos \cite{vi2}-\cite{vi3}, the vector inflation model proposed by Golovenev, Mukhanov, and Vanchurin \cite{Golovnev:2008cf}, and the hybrid inflation model with coupled vector and scalar ``waterfall field'' where $\zeta$ is generated at the end of inflation proposed by Yokoyama and Soda \cite{vi} (see also an interesting critical discussion about this work in \cite{varios8}).  The latter results sounds ubiquitous for any variety of the SY consistency relation, that does not involve integrations over momentum space, when there is no scale-invariance; nonetheless, we have been able to introduce the fifth variety of the SY consistency relation in Eq. (\ref{5vsy}) whose validity is only restricted to the existence of statistical homogeneity and when $f_{\rm NL} ({\bf k}_1, {\bf k}_2, {\bf k}_3)$ and $\tau_{\rm NL} ({\bf k}_1, {\bf k}_2, {\bf k}_3, {\bf k}_4)$ are calculated in the squeezed and collapsed limits respectively; indeed, this variety stands the test of the class of models described above as shown in Section \ref{fourth};  perhaps the strongest conclusion we can extract from all this analysis is a simple question:  what will we do if the fifth variety of the SY consistency relation is observationally violated?  It would imply statistical inhomogeneity, which in turn implies that, although we dispose of very good data and very good theory, we cannot compare them to each other \cite{ValenzuelaToledo:2011fj,weinbergbook}:  this situation would be a real shake in the foundations of cosmology as a science \cite{yrodrig}.

\section*{Acknowledgments}
Y.R. and J.P.B.A. are supported by Fundaci\'on para la Promoci\'on de la Investigaci\'on y la Tecnolog\'{\i}a del Banco de la Rep\'ublica (COLOMBIA) grant number 3025 CT-2012-02 and by VCTI (UAN) grant number 2011254. Y.R. was a JSPS postdoctoral fellow (P11323) and, in addition, is supported by DIEF de Ciencias (UIS) grant number 5177.  C.A.V.-T. is supported by Vicerrector\'{\i}a de Investigaciones (UNIVALLE) grant number 7858.  J.P.B.A. acknowledges the warm hospitality of Departamento de F\'{\i}sica at Universidad del Valle where part of this work was carried out.


\end{document}